\documentclass[a4paper,11pt]{article}

\usepackage{jheppub}

\usepackage{amsfonts,graphics,epsfig,subfigure}

\title{Ratio of critical quantities related to Hawking temperature-entanglement entropy criticality}

\author[]{Jie-Xiong Mo,}
\author[]{Gu-Qiang Li}

 \affiliation[]{Institute of Theoretical Physics, Lingnan Normal University, Zhanjiang, 524048, Guangdong, China}

\emailAdd{mojiexiong@gmail.com}
\emailAdd{zsgqli@hotmail.com}

\abstract{We revisit the Hawking temperature$-$entanglement entropy criticality of the $d$-dimensional charged AdS black hole with our attention concentrated on the ratio $\frac{T_c \delta S_c}{Q_c}$. Comparing the results of this paper with those of the ratio $\frac{T_c S_c}{Q_c}$, one can find both the similarities and differences. These two ratios are independent of the characteristic length scale $l$ and dependent on the dimension $d$. These similarities further enhance the relation between the entanglement entropy and the Bekenstein-Hawking entropy. However, the ratio $\frac{T_c \delta S_c}{Q_c}$ also relies on the size of the spherical entangling region. Moreover, these two ratios take different values even under the same choices of parameters. The differences between these two ratios can be attributed to the peculiar property of the entanglement entropy since the research in this paper is far from the regime where the behavior of the entanglement entropy is dominated by the thermal entropy.}

\begin{document}
\maketitle
\flushbottom

\section{Introduction}

Entanglement entropy has received considerable attention these years for its important position in the AdS/CFT correspondence and widespread application in probing various physical phenomena. The Ryu-Takayanagi formula \cite{Takayanagi1,Takayanagi2,Takayanagi3} shares striking similarity with the Bekenstein-Hawking entropy, implying that there may exist close relation between the entanglement entropy and black hole entropy. This relation sheds light on some deep physics. It was proposed that the entanglement entropy is the origin of black hole entropy \cite{Srednicki,Frolov,Solodukhin}. Recently, Johnson \cite{Johnson} further enhanced this relation by disclosing intriguing phase structures of entanglement entropy. The isocharges in the entanglement entropy-temperature plane exhibit similarly to those of the thermal entropy-temperature plane. Moreover, it was shown that they share the same critical temperature and critical exponents \cite{Johnson}. Soon afterwards, the equal area law was proved to hold for the entanglement entropy-temperature plane \cite{Nguyen}, just as it holds for $T-S$ (Here, $S$ denotes the thermal entropy of the black hole.) curve \cite{Spallucci}. Ever since the pioneer work \cite{Johnson} of Johnson, the rich phase structures of various black holes have been investigated from the perspective of entanglement entropy \cite{Caceres}-\cite{zengxiaoxiong5}.

In our recent paper \cite{xiong10}, we concentrate on the ratios of critical physical quantities related to three different kinds of criticality of $d$-dimensional charged AdS black holes. Namely, $T-S$ criticality, $Q-\Phi$ criticality and $P-V$ criticality. In all these cases, we showed that there exist universal ratios that do not depend on the parameters. We also showed that the value of $\frac{T_cS_c}{Q_c}$ for $T-S$ criticality differs from that of $\frac{P_cv_c}{T_c}$ for $P-V$ criticality. Probing the universal ratios of critical quantities is of great physical significance. Disclosing the deep physics behind the phenomena of universal ratios will help draw a unified picture of black hole thermodynamics.

     Considering the close relation between the entanglement entropy and the Bekenstein-Hawking entropy (as stated in the first paragragh), there may exist universal ratio of critical quantities related to the Hawking temperature-entanglement entropy criticality. This issue has not been covered in literature yet to the best of our knowledge. And investigating this ratio will be the target of this paper. We are about to resolve this issue within the framework of $d$-dimensional charged AdS black hole spacetime. The motivations are as follows. Firstly, we are curious about whether the analogous ratio of critical quantities related to the Hawking temperature-entanglement entropy criticality is also universal. In other words, does it depend on the parameters? Probing the universal ratios of critical quantities of charged AdS black holes has its own right. The analogy between charged AdS black holes and van der Waals liquid-gas system has gained extensive attention ever since the famous work \cite{Chamblin1,Chamblin2}. The ratio $\frac{P_cv_c}{kT_c}$ is a universal number for all van der Waals fluids in classical thermodynamics, motivating us to searching for the universal ratios for charged AdS black holes. Secondly, we are interested in both the similarities and differences (if any) between the ratio for the Hawking temperature-entanglement entropy criticality and that for the Hawking temperature-thermal entropy criticality. On the one hand, the similarities will help further understand the close relation between the entanglement entropy and the Bekenstein-Hawking entropy. On the other hand, the differences may shed light on some yet unknown physics.

    The organization of this paper is as follows. Sec.\ref{Sec2} devotes to a short review of three different kinds of criticality of $d$-dimensional charged AdS black holes. In Sec.\ref{Sec3}, we will revisit the Hawking temperature$-$entanglement entropy criticality of the $d$-dimensional charged AdS black hole and investigate the analogous ratio of critical quantities for various cases where different parameters are chosen as the variable respectively. In the end, Sec.\ref{Sec4} devotes to conclusions.

\section{A short review of criticality of charged AdS black holes}
\label{Sec2}

The metric of the $d$-dimensional ($d>3$) charged AdS black hole reads
\begin{equation}
ds^2=-f(r)dt^2+\frac{dr^2}{f(r)}+r^2d\Omega_{d-2}^2,\label{1}
\end{equation}%
where
\begin{equation}
f(r)=1-\frac{m}{r^{d-3}}+\frac{q^2}{r^{2(d-3)}}+\frac{r^2}{l^2}.\label{2}
\end{equation}%
$l$ is the characteristic length scale which is related to the cosmological constant $\Lambda$ through $\Lambda=-\frac{(d-1)(d-2)}{2l^2}$. Parameters $m$ and $q$ can be identified with the ADM mass and the electric charge of the black hole as follows~\cite{Chamblin1}
\begin{eqnarray}
M&=&\frac{\omega_{d-2}(d-2)}{16\pi}m,\label{3}
\\
Q&=&\frac{\omega_{d-2}\sqrt{2(d-2)(d-3)}}{8\pi}q,\label{4}
\end{eqnarray}%
where the volume of the unit $d$-sphere $\omega_d$ can be obtained via $\frac{2\pi^{\frac{d+1}{2}}}{\Gamma(\frac{d+1}{2})}$  .

The Hawking temperature, the entropy and the electric potential of the $d$-dimensional ($d>3$) charged AdS black hole have been reviewed in Ref.~\cite{Gunasekaran} as
\begin{eqnarray}
T&=&\frac{f'(r_+)}{4\pi}=\frac{d-3}{4\pi r_+}\left(1-\frac{q^2}{r_+^{2(d-3)}}+\frac{d-1}{d-3} \frac{r_+^2}{l^2}\right),\label{5}
\\
S&=&\frac{\omega_{d-2}r_+^{d-2}}{4}, \label{6}
\\
\Phi&=&\sqrt{\frac{d-2}{2(d-3)}}\frac{q}{r_+^{d-3}}. \label{7}
\end{eqnarray}%

Ref.~\cite{Gunasekaran} also investigated its $P-V$ criticality and obtained the following critical quantities
\begin{eqnarray}
v_c&=&\frac{1}{\kappa}\left[q^2(d-2)(2d-5)\right]^{1/[2(d-3)]},\label{8} \\
T_c&=&\frac{(d-3)^2}{\pi \kappa v_c(2d-5)},\label{9} \\
P_c&=&\frac{(d-3)^2}{16\pi \kappa^2 v_c^2},\label{10}
\end{eqnarray}%
where $\kappa=\frac{d-2}{4}$. Note that the cosmological constant has been identified as the thermodynamic pressure through the definition $P=-\Lambda/(8\pi)$ and the specific volume $v$ is related to the horizon radius $r_+$ through $r_+=\kappa v$. The ratio $\frac{P_cv_c}{T_c}$ was shown to be $\frac{2d-5}{4d-8}$, which is independent of the parameter $q$~\cite{Gunasekaran}.

In our recent work~\cite{xiong10}, we studied the $T-S$ criticality of these black holes and probed the possible universal ratio for $T-S$ criticality. Here, we listed the results for the $d$-dimensional ($d>3$) charged AdS black hole.
\begin{eqnarray}
S_c&=&\frac{\omega_{d-2}}{4}\left[\frac{(d-3)l}{\sqrt{(d-2)(d-1)}}\right]^{d-2},\label{11} \\
Q_c&=&\frac{(d-3)^{\frac{2d-5}{2}}l^{d-3}\omega_{d-2}}{4\sqrt{4d-10}[(d-2)(d-1)]^{\frac{d-3}{2}}\pi},\label{12} \\
T_c&=&\frac{(d-3)\sqrt{(d-2)(d-1)}}{(2d-5)l\pi},\label{13} \\
\frac{T_cS_c}{Q_c}&=&\frac{\sqrt{2}(d-3)^{3/2}}{\sqrt{2d-5}}.\label{14}
\end{eqnarray}%
Note that the ratio $\frac{T_cS_c}{Q_c}$ ($S_c$ denotes the critical thermal entropy) is independent of the parameter $l$ although the relevant critical quantities $S_c, Q_c, T_c$ all depend on $l$. In this sense, the ratio $\frac{T_cS_c}{Q_c}$ can be viewed as a universal ratio for the $T-S$ criticality. And the value of this ratio differs from that of $\frac{P_cv_c}{T_c}$.

Ref.~\cite{mayubo} studied $Q-\Phi$ criticality of these black holes and obtained
\begin{eqnarray}
\Phi_c&=&\frac{\pi^{n/2-1}}{4\Gamma(n/2)}\sqrt{\frac{n-1}{2n-3}},\label{15} \\
T_c&=&\frac{n-2}{\pi (2n-3)}(-2\Lambda)^{1/2},\label{16} \\
Q_c&=&[-\frac{(n-2)^2}{2\Lambda}]^{(n-2)/2}[(n-1)(2n-3)]^{-1/2},\label{17} \\
\frac{\Phi_cQ_c}{T_c}&=&\frac{\sqrt{-\Lambda}2^{-\frac{n}{2}-\frac{3}{2}}(n-1)\pi^{n/2}[-\frac{(n-2)^2}{\Lambda}]^{n/2}}{(n-2)^3\sqrt{\frac{n-1}{2n-3}}\sqrt{2n^2-5n+3}\Gamma\left(\frac{n}{2}\right)},\label{18}
\end{eqnarray}%
where $n=d-1$. They argued that the ratio $\frac{\Phi_cQ_c}{T_c}$ depends on both $\Lambda$ and $n$ and is not universal. In our recent work~\cite{xiong10}, we construct two ratios for the $Q-\Phi$ criticality as follows
\begin{eqnarray}
\frac{\Phi_cQ_c^{\frac{1}{2-n}}}{T_c}&=&\frac{[(2n-3)(n-1)]^{\frac{n-1}{2(n-2)}}\pi^{n/2}}{4(n-2)^2\Gamma\left(\frac{n}{2}\right)},\label{19} \\
\frac{\Phi_cQ_c}{T_c^{2-n}}&=&\frac{(n-2)^{2n-4}(2n-3)^{1-n}\pi^{1-n/2}}{4\Gamma\left(\frac{n}{2}\right)}.\label{20}
\end{eqnarray}%
Note that these two ratios only depend on $n$.

\section{Ratio of critical quantities for Hawking temperature$-$entanglement entropy criticality}
\label{Sec3}
 In the former section, we review the ratios of critical quantities for three different kinds of criticality. Namely, $P-V$ criticality, $T-S$ criticality and $Q-\Phi$ criticality. In all these cases, there exist universal ratios. Considering the close relation between the entanglement entropy and the Bekenstein-Hawking entropy, we will revisit the Hawking temperature$-$entanglement entropy criticality of the $d$-dimensional charged AdS black hole and probe the analogous ratio of critical quantities.

 Suppose $\Sigma$ is the codimension-2 minimal surface with boundary condition $\partial \Sigma=\partial A$, the entanglement entropy $S_A$ between the region $A$ and its complement can be defined holographically as \cite{Takayanagi1,Takayanagi2,Takayanagi3}
 \begin{equation}
S_A=\frac{Area(\Sigma)}{4G_N},\label{21}
\end{equation}
where $G_N$ is the Newton's constant and $Area(\Sigma)$ denotes the area of $\Sigma$ (a minimal surface anchored on $\partial A$).

To avoid dealing with the phase transition between connected and disconnected minimal surfaces, one can consider a spherical cap on the boundary delimited by $\theta\leq\theta_0$ as Ref. \cite{Nguyen} did. Then the minimal surface can be parametrized by $r(\theta)$. Utilizing Eq.(\ref{21}), the entanglement entropy can be derived as
 \begin{equation}
S_A=\frac{\pi}{2}\int^{\theta_0}_0 r\sin \theta \sqrt{\frac{r'^2}{f(r)}+r^2} d \theta,\label{22}
\end{equation}
where $r'=dr/ d\theta$. Note that $r(\theta)$ can be obtained by solving the famous Euler-Lagrange equation
\begin{equation}
\frac{\partial \mathcal{L}}{\partial r}=\frac{d}{d\theta}\left(\frac{\partial \mathcal{L}}{\partial r'}\right), \label{23}
\end{equation}
where $\mathcal{L}$ can be read from Eq.(\ref{22}) and the boundary condition can be chosen as $r(0)= r_0, r'(0)=0$.

With the numerical solution of $r(\theta)$, one can calculate the holographic entanglement entropy by utilizing Eq.(\ref{22}). Note that the result should be regularized by subtracting the entanglement entropy in pure AdS $S_0$ with the same boundary region to avoid the divergence. And the regularized entanglement entropy $\delta S$ reads $S_A-S_0$. The analytic result for $r_{AdS}(\theta)$ corresponding to $S_0$ was presented as \cite{Blanco, Casini}
\begin{equation}
r_{AdS}(\theta)=l \left[\left(\frac{\cos \theta}{\cos \theta_0}\right)^2-1\right]^{-1/2}. \label{24}
\end{equation}

Here, we are interested in the ratio $\frac{T_c \delta S_c}{Q_c}$, which is analogous to the ratio $\frac{T_cS_c}{Q_c}$. Note that $\delta S_c$ denotes the regularized entanglement entropy at the critical point. Specifically, we will study the cases where $l$, $d$ and $\theta_0$ are chosen as the variable respectively to probe whether the ratio $\frac{T_c \delta S_c}{Q_c}$ is universal.

\begin{figure*}
\centerline{\subfigure[]{\label{1a}
\includegraphics[width=8cm,height=6cm]{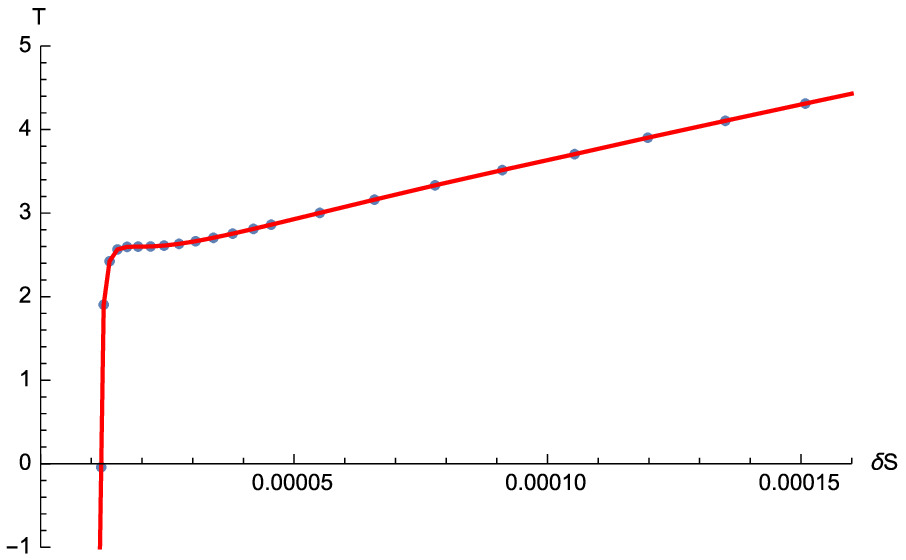}}
\subfigure[]{\label{1b}
\includegraphics[width=8cm,height=6cm]{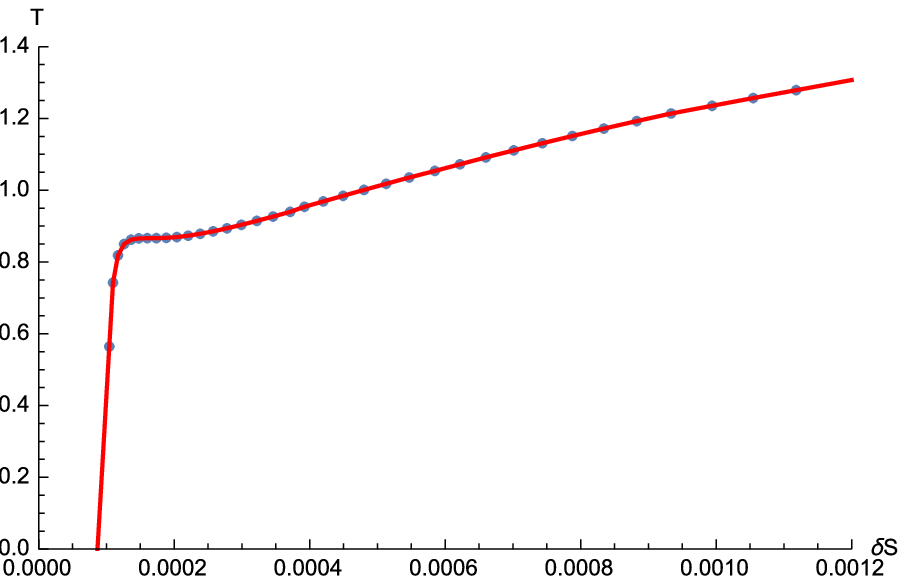}}}
\centerline{\subfigure[]{\label{1c}
\includegraphics[width=8cm,height=6cm]{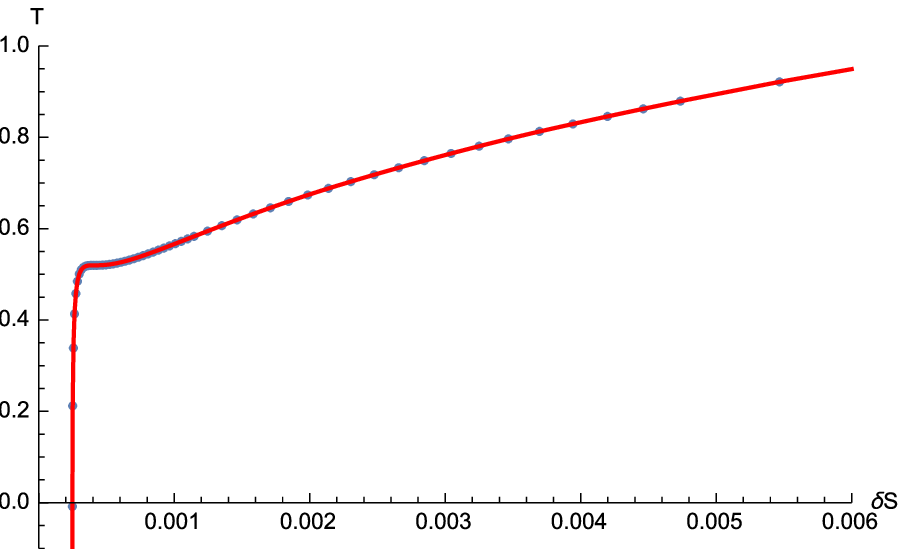}}
\subfigure[]{\label{1d}
\includegraphics[width=8cm,height=6cm]{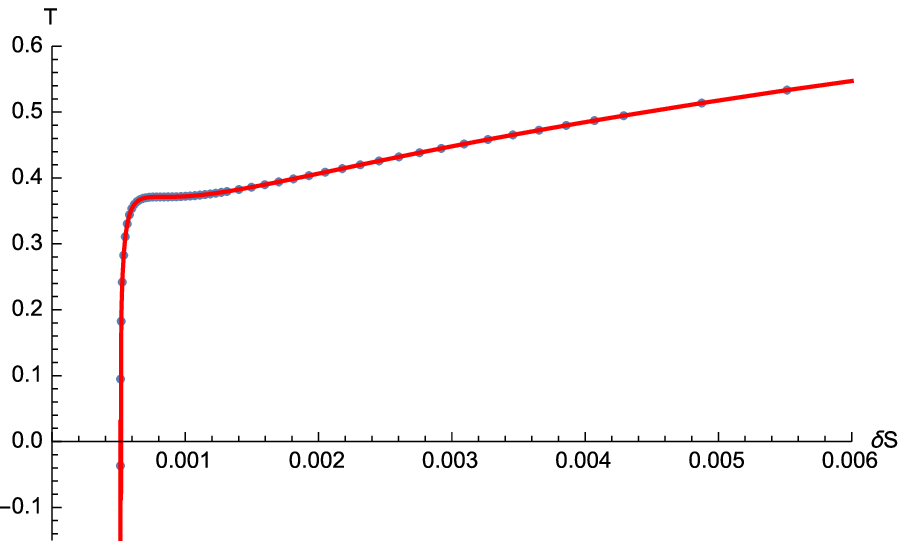}}}
\centerline{\subfigure[]{\label{1e}
\includegraphics[width=8cm,height=6cm]{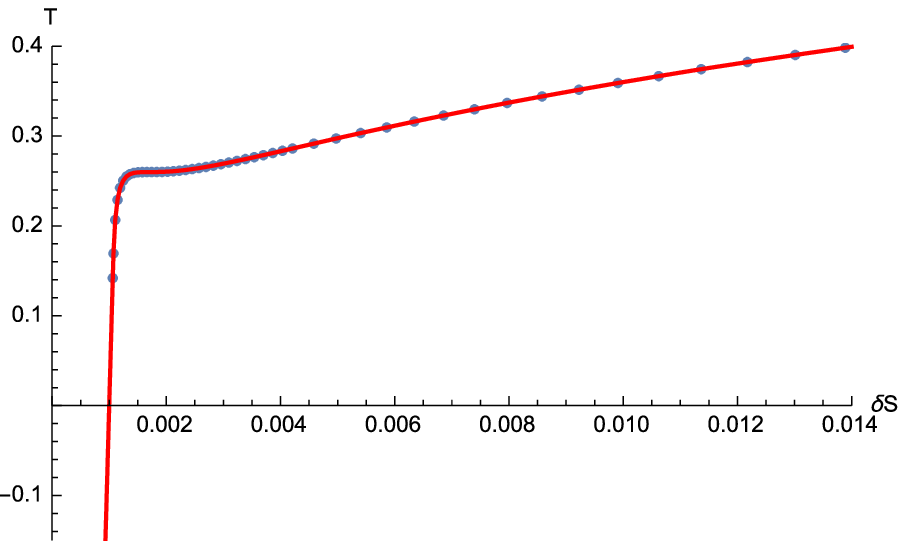}}
\subfigure[]{\label{1f}
\includegraphics[width=8cm,height=6cm]{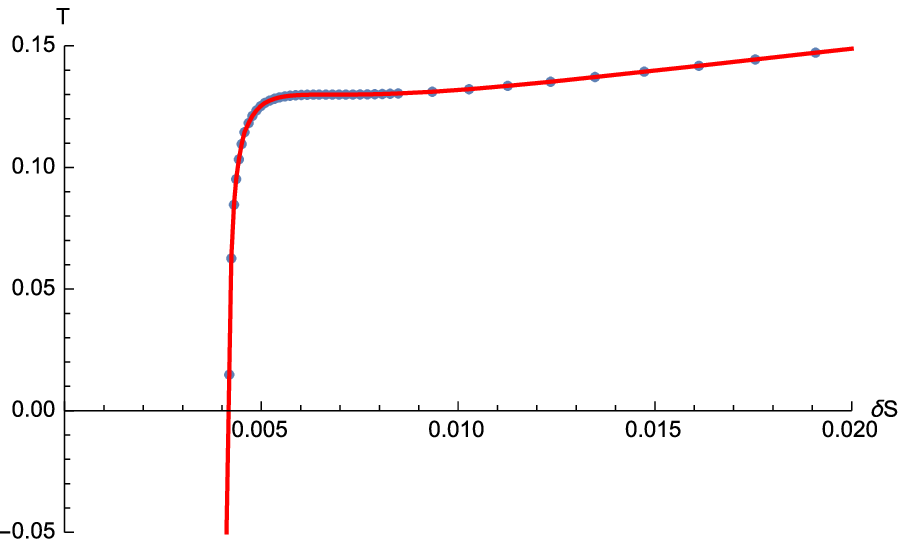}}}
 \caption{$T$ vs. $\delta S$ for $d=4, \theta_0=0.2, Q=Q_c$ (a) $l=0.1$ (b) $l=0.3$ (c) $l=0.5$ (d) $l=0.7$ (e) $l=1$ (f) $l=2$} \label{fg1}
\end{figure*}

Firstly, we fix $d=4$ and $\theta_0=0.2$ and let $l$ vary from $0.1$ to $2$. The cutoff $\theta_c$ is chosen as 0.199. Since the focus of our research is the critical quantities, we focus on the case $Q=Q_c$ and ignore the cases $Q<Q_c$ and $Q>Q_c$. The $T-\delta S$ curves corresponding to different choices of $l$ are depicted in Fig. \ref{1a}-\ref{1f} while the relevant critical physical quantities are listed in Table \ref{tb1}. With the increasing of $l$, the critical quantities $Q_c$, $S_c$ and $\delta S_c$ increase while $T_c$ decreases. The ratio $\frac{T_cS_c}{Q_c}$ is almost unchanged, just as the analytic result we obtained before showed~\cite{xiong10}. The ratio $\frac{T_c \delta S_c}{Q_c}$ changes slightly. The mean value is $0.00271188$ while the standard deviation turns out to be $0.00015822$. The standard deviation is so small that one can also conclude that the ratio $\frac{T_c \delta S_c}{Q_c}$ does not depend on the characteristic length scale $l$.

\begin{table}[!h]
\tabcolsep 0pt
\caption{Critical physical quantities for $d=4,\theta_0=0.2$}
\vspace*{-12pt}
\begin{center}
\def\temptablewidth{1\textwidth}
{\rule{\temptablewidth}{2pt}}
\begin{tabular*}{\temptablewidth}{@{\extracolsep{\fill}}cccccccc}
$l$ & $Q_c$ & $T_c$ &$S_c$ &$\delta S_c$ &$\frac{T_cS_c}{Q_c}$ & $\frac{T_c\delta S_c}{Q_c}$  \\   \hline
    0.1  & 0.01666667     & 2.59898934        &        0.00523599 &0.00001923& 0.81649677 & 0.00299871  \\
    0.3    & 0.05     & 0.86632978       &      0.04712389 & 0.00016070 & 0.81649659 & 0.00278439  \\
    0.5    & 0.08333333    & 0.51979787       &      0.13089969 & 0.00041065 & 0.81649659 & 0.00256146  \\
0.7     &  0.11666667    & 0.37128419        &        0.25656340 & 0.00083277 & 0.81649655 & 0.00265024 \\
 1     &  0.16666667    & 0.25989893       &         0.52359877 & 0.00168325 & 0.81649654 & 0.00262484 \\
 2     &  0.33333333   & 0.12994946       &       2.09439510 & 0.00680175 & 0.81649654 & 0.00265165
       \end{tabular*}
       {\rule{\temptablewidth}{2pt}}
       \end{center}
       \label{tb1}
       \end{table}

\begin{figure*}
\centerline{\subfigure[]{\label{2a}
\includegraphics[width=8cm,height=6cm]{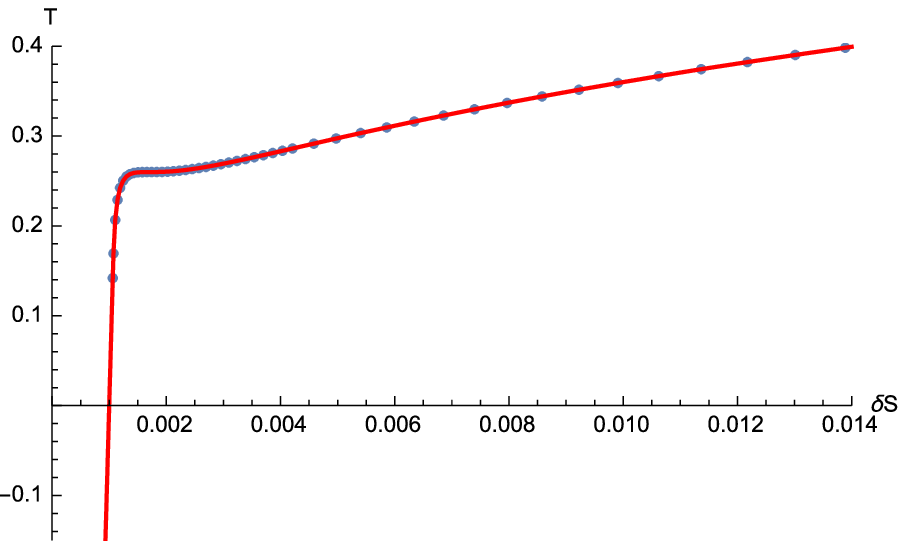}}
\subfigure[]{\label{2b}
\includegraphics[width=8cm,height=6cm]{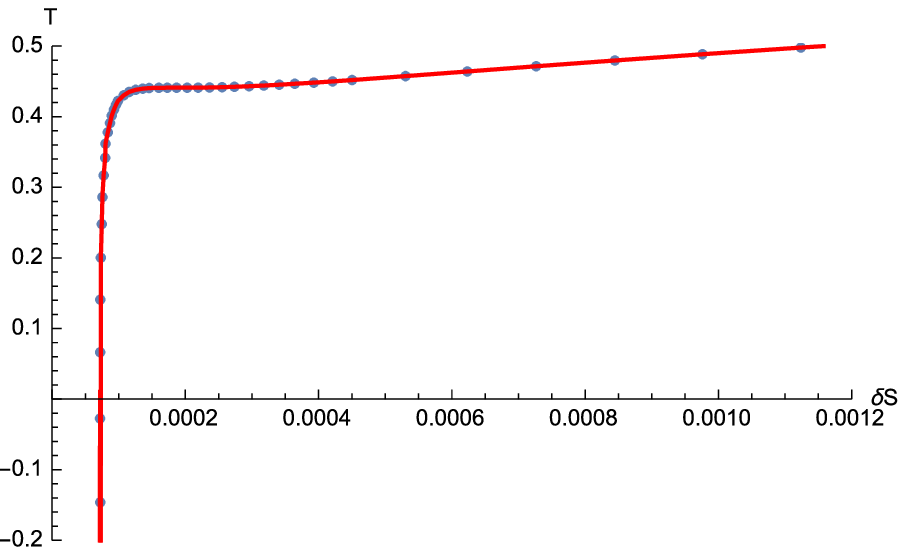}}}
\centerline{\subfigure[]{\label{2c}
\includegraphics[width=8cm,height=6cm]{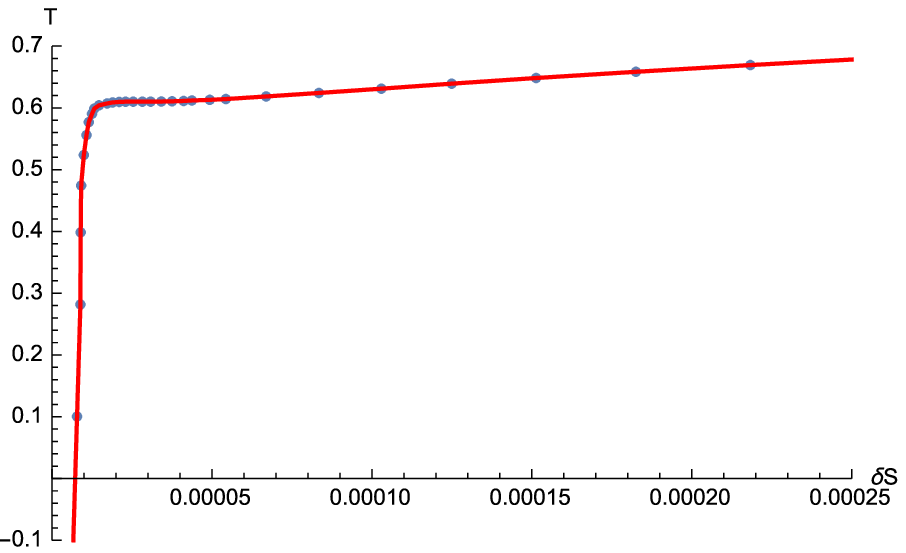}}
\subfigure[]{\label{2d}
\includegraphics[width=8cm,height=6cm]{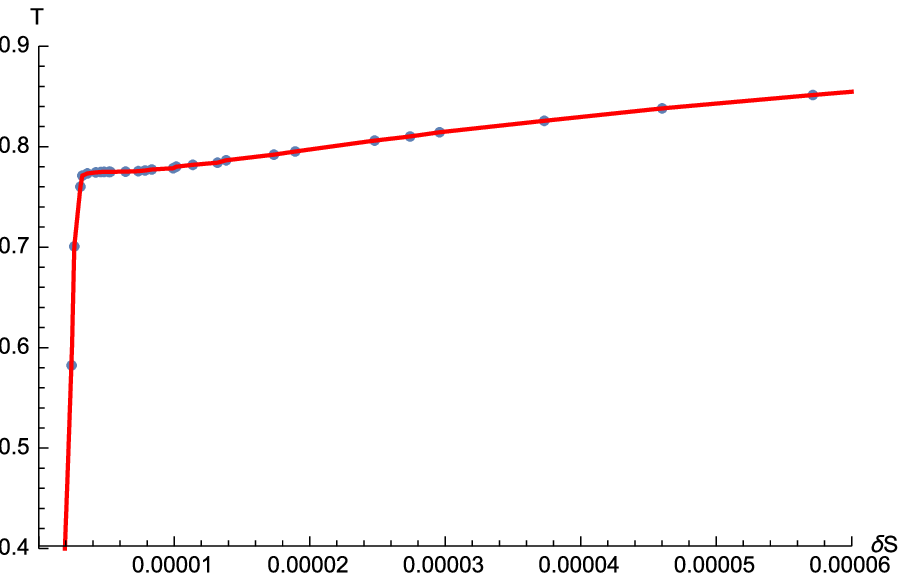}}}
\centerline{\subfigure[]{\label{2e}
\includegraphics[width=8cm,height=6cm]{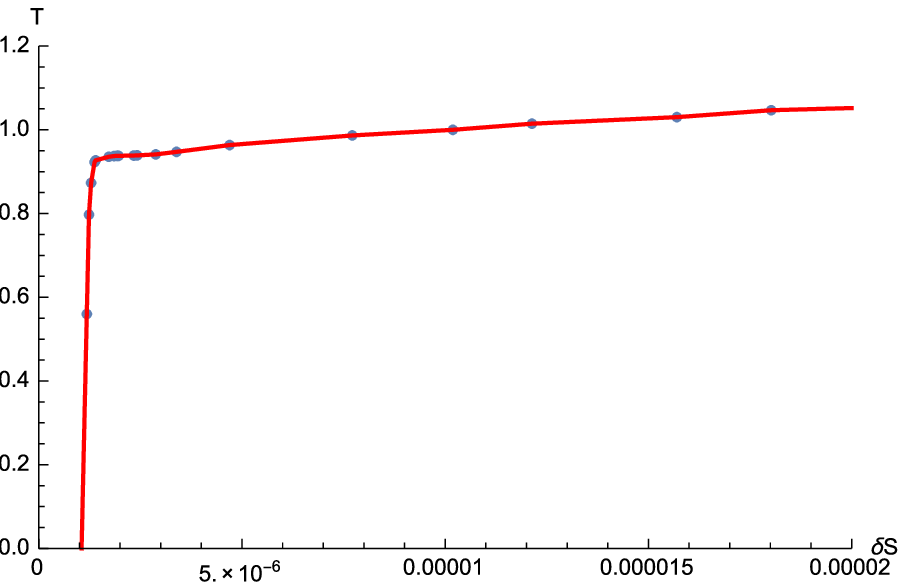}}}
 \caption{$T$ vs. $\delta S$ for $l=1, \theta_0=0.2, Q=Q_c$ (a) $d=4$ (b) $d=5$ (c) $d=6$ (d) $d=7$ (e) $d=8$} \label{fg2}
\end{figure*}

Secondly, we fix $l=1$ and $\theta_0=0.2$ and let $d$ vary from $4$ to $8$. The cutoff $\theta_c$ is chosen as 0.199. Fig. \ref{2a}-\ref{2e} show the corresponding $T-\delta S$ curves for different $d$. And the relevant critical physical quantities are listed in Table \ref{tb2}. It can be clearly witnessed that both the ratio $\frac{T_c S_c}{Q_c}$ and $\frac{T_c \delta S_c}{Q_c}$ change with $n$. With the increasing of $n$, the ratio $\frac{T_c S_c}{Q_c}$ increases while the ratio $\frac{T_c \delta S_c}{Q_c}$ decreases. So these ratios are dimensionality dependent.

\begin{table}[!h]
\tabcolsep 0pt
\caption{Critical physical quantities for $l=1,\theta_0=0.2$}
\vspace*{-12pt}
\begin{center}
\def\temptablewidth{1\textwidth}
{\rule{\temptablewidth}{2pt}}
\begin{tabular*}{\temptablewidth}{@{\extracolsep{\fill}}cccccccc}
$d$ & $Q_c$ & $T_c$ &$S_c$ &$\delta S_c$ &$\frac{T_cS_c}{Q_c}$ & $\frac{T_c\delta S_c}{Q_c}$  \\   \hline
   4     &  0.16666667    & 0.25989893       &         0.52359877 & 0.00168325 & 0.81649654 & 0.00262484 \\
5    &  0.23416049     & 0.44106312       &        0.94970312 &0.00018683 & 1.78885441 & 0.00035191 \\
 6   &  0.29266716     & 0.61008218        &      1.33239659 & 0.00002721 & 2.77746029 & 0.00005672 \\
 7   &  0.33084986    & 0.77486891       &      1.61022456 & $5.02138069\times10^{-6}$ & 3.77123614 &0.00001176 \\
  8   & 0.34298011     & 0.93767447       &        1.74377523 & $1.51717870\times10^{-6}$ & 4.76731293 & $4.14781992\times10^{-6}$
       \end{tabular*}
       {\rule{\temptablewidth}{2pt}}
       \end{center}
       \label{tb2}
       \end{table}

\begin{figure*}
\centerline{\subfigure[]{\label{3a}
\includegraphics[width=8cm,height=6cm]{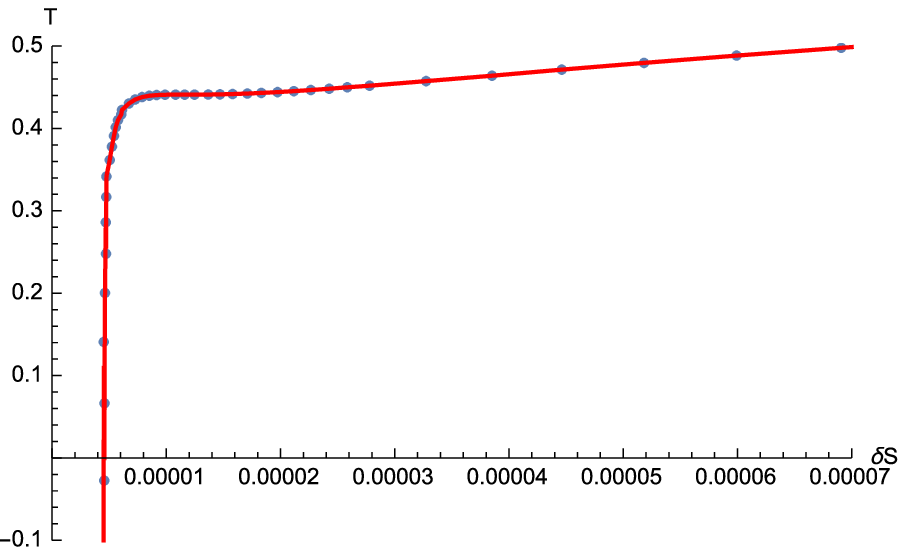}}
\subfigure[]{\label{3b}
\includegraphics[width=8cm,height=6cm]{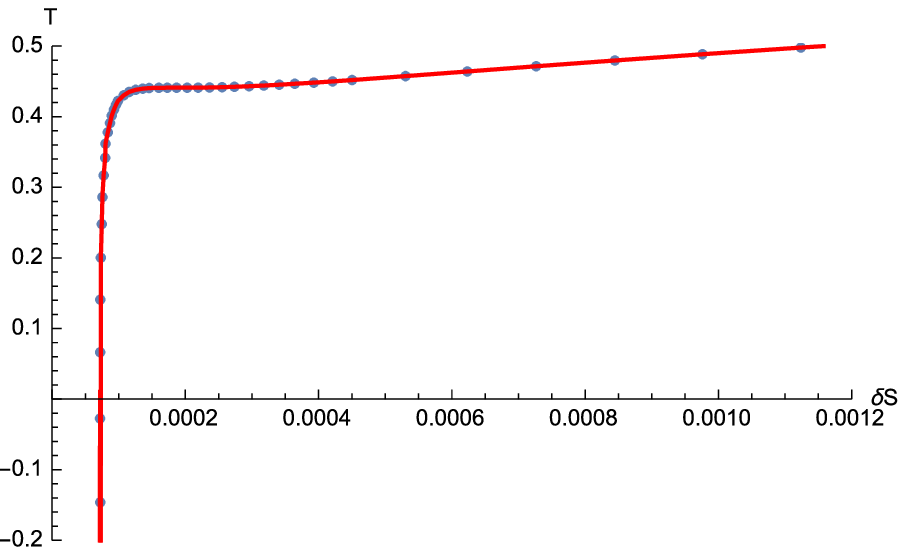}}}
\centerline{\subfigure[]{\label{3c}
\includegraphics[width=8cm,height=6cm]{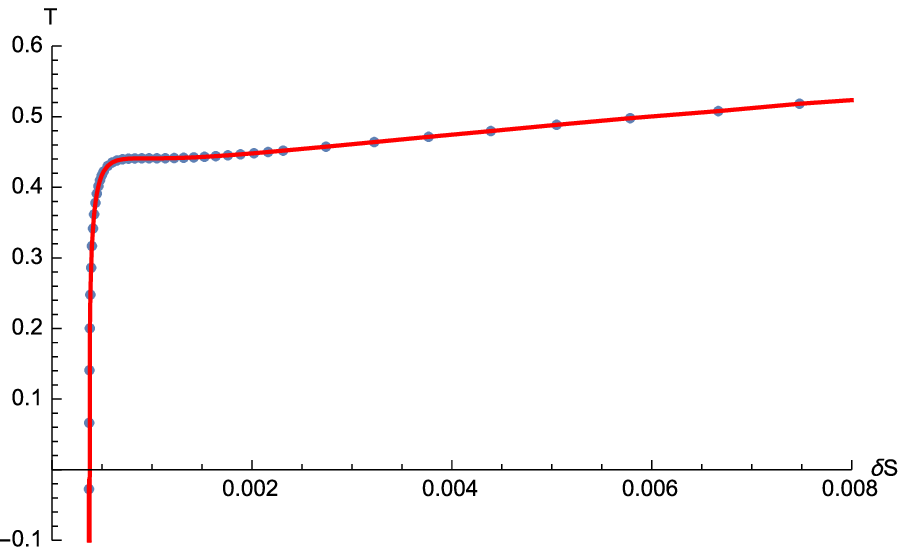}}
\subfigure[]{\label{3d}
\includegraphics[width=8cm,height=6cm]{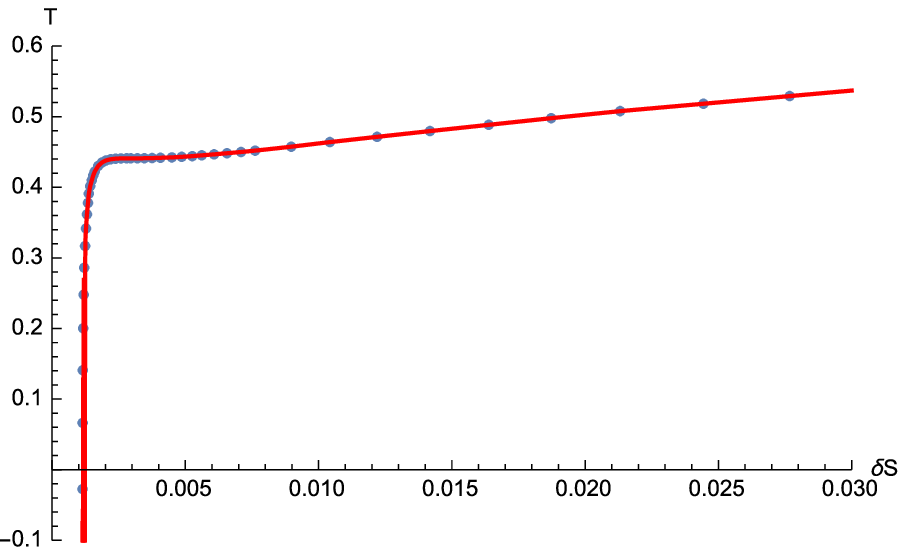}}}
 \caption{$T$ vs. $\delta S$ for $d=5, l=1, Q=Q_c$ (a) $\theta_0=0.1$ (b) $\theta_0=0.2$  (c) $\theta_0=0.3$ (d) $\theta_0=0.4$} \label{fg3}
\end{figure*}

Thirdly, we fix $l=1$ and $d=5$ and let $\theta_0$ vary from $0.1$ to $0.4$. The cutoff $\theta_c$ is chosen as $0.099, 0.199, 0.299, 0.399$ respectively. The corresponding $T-\delta S$ curves for different choices of $\theta_0$ are shown in Fig. \ref{3a}-\ref{3d} while the relevant critical physical quantities are listed in Table \ref{tb3}. With the increasing of $\theta_0$, the ratio $\frac{T_c S_c}{Q_c}$ remains unchanged while $\frac{T_c \delta S_c}{Q_c}$ increases. It is not difficult to explain this result considering the observation that $\delta S_c$ increases with $\theta_0$ while $S_c$ is not affected.

 \begin{table}[!h]
\tabcolsep 0pt
\caption{Critical physical quantities for $d=5,l=1$}
\vspace*{-12pt}
\begin{center}
\def\temptablewidth{1\textwidth}
{\rule{\temptablewidth}{2pt}}
\begin{tabular*}{\temptablewidth}{@{\extracolsep{\fill}}cccccccc}
$\theta_0$ & $Q_c$ & $T_c$ &$S_c$ &$\delta S_c$ &$\frac{T_cS_c}{Q_c}$ & $\frac{T_c\delta S_c}{Q_c}$  \\   \hline
  0.1   &  0.23416049     & 0.44106312       &        0.94970312 & 0.00001161 & 1.78885441 & 0.00002187 \\
0.2   &  0.23416049     & 0.44106312       &        0.94970312 &0.00018683 & 1.78885441 & 0.00035191 \\
0.3   &  0.23416049     & 0.44106312       &        0.94970312 &0.00097004& 1.78885441 & 0.00182716  \\
0.4   &  0.23416049     & 0.44106312       &        0.94970312 &0.00295175& 1.78885441 & 0.00555990
       \end{tabular*}
       {\rule{\temptablewidth}{2pt}}
       \end{center}
       \label{tb3}
       \end{table}

\section{Concluding Remarks}
\label{Sec4}
    To summarize, we revisit the Hawking temperature$-$entanglement entropy criticality of the $d$-dimensional charged AdS black hole and concentrate our attention on the ratio of critical quantities. Specifically, we calculated numerically the ratio $\frac{T_c \delta S_c}{Q_c}$ for the cases where $l$, $d$ and $\theta_0$ are chosen as the variable respectively.

Comparing the results of this paper with those of the ratio $\frac{T_c S_c}{Q_c}$ \cite{xiong10}, one can find that both the similarities and differences exist. These two ratios are independent of the characteristic length scale $l$ and dependent on the dimension $d$. These similarities further enhance the relation between the entanglement entropy and the Bekenstein-Hawking entropy.

Contrary to the ratio $\frac{T_c S_c}{Q_c}$, the ratio $\frac{T_c \delta S_c}{Q_c}$ relies on the size of the spherical entangling region. Moreover, these two ratios take different values under the same choices of parameters. Note that we focus on the entanglement for a subsystem whose volume is very small. In this sense, the research in this paper is far from the regime where the behavior of the entanglement entropy is dominated by the thermal entropy. So the differences between these two ratios may be attributed to the peculiar property of the entanglement entropy. And the deep physics behind it certainly deserves more attention in the future research.

\acknowledgments The authors are supported by National Natural Science Foundation of China (Grant No.11605082), and in part supported by Natural Science Foundation of Guangdong Province, China (Grant Nos.2016A030310363, 2016A030307051, 2015A030313789).

\end{document}